**Unpublished Opening Lecture for the Course on the Theory of Relativity in Argentina, 1925[*]**

*Albert Einstein*

Translated by Alejandro Gangui and Eduardo L. Ortiz

Honorable Rector, Honorable Professors, and Students of this University: In these times of political and economic struggle and nationalistic fragmentation, it is a particular joy for me to see people assembling here to give their attention exclusively[1] to the highest values that are common to us all. I am glad to be in this blessed land before a small circle of people who are interested in topics of science to speak on those issues that, in essence, are the subject of my own meditations.

In science, there are always two opposite and complementary goals that, owing to their mutual complementarity, constitute the essence of its progress. On the one hand, there is the quest for enlargement[2] and enrichment of our understanding of some particular area of knowledge; and on the other hand, there is the endeavor to achieve a systematic unity of knowledge. In my work I have always attempted the latter; therefore, I wish to communicate here more accurate observations on this goal, the systematic unity of knowledge.

Using as few hypothetical laws as possible, science attempts to explain relations between observable facts, arriving at them in a deductive manner, that is, in a purely logical way. Physics is customarily referred to as an empirical science and it is believed that its fundamental laws are deduced from experiments, so as to indicate how it differs from speculative philosophy. However, in truth the relationship between fundamental laws and facts from experience is not that simple. Indeed, there is no scientific method to deduce inductively these fundamental laws from experimental data. The formulation of a fundamental law is, rather, an act of intuition which can be achieved only by one who watches empirically with the necessary attention and has sufficient empirical understanding of the field in question. The sole criteria for the truth of a fundamental law is only that we can be sure that the relations between observable events can be logically deduced from it. It follows then that a fundamental law can be refuted in a definite manner, but can never be definitely shown to be correct, as one must always bear in mind the possibility of discovering a new phenomenon that contradicts the logical conclusions arising from a fundamental law.

---

[1] Einstein added the word "exclusively" to his original draft.
[2] The text to this point is translated from the first page of the German manuscript of the lecture in Einstein's handwriting. It was reproduced photographically and appeared in *La Vida Literaria*, Buenos Aires, 1931 (see fig. 2 in the Appendix) as well as in J. A. Stargardt's catalogs 615 (1978), 117, and 683 (2006), 189 (see fig. 3 in the Appendix). The text from this point ("und Bereicherung unseres Einzel-Wissens ..." ) to footnote 3 ("... durch eine zwangläufige Methode zu ersetzen ..."), is translated from the German excerpt printed in J. A. Stargardt's catalog 683 (2006), 188 (see fig. 4 in the Appendix).

Experience is, therefore, the judge, but not the generator of fundamental laws. The transition from the facts of experience to a fundamental law often requires an act of free creativity from our imagination, as well as an act of creation of concepts and relations; it would not be possible to replace this act with a necessary and conclusive method.[3]

The fact that a concept in the presence of experience, even if originated from experience, has a certain logical independence is appreciated by considering extra-scientific thought. The observation of the existence of similar objects has given rise to the notion of number, but has not created it. In fact, people in some cultures have not gone any further than an understanding of only the smallest of numbers.

Returning to the ideas and fundamental laws of physics, it is easy to show that starting from the facts of experience there is no fixed road taking us back to those ideas and fundamental laws. Let us consider, for example, the laws of motion on which classical astronomy rests. Using logical and mathematical methods we can deduce from Kepler's laws Newton's law on the [inverse][4] proportionality of force[5] on the square of the distances. But

---

[3] The text from here to footnote 5, is translated from Spanish, which was published in *La Vida Literaria*.
[4] The word in square brackets is added by the translators.

Galileo's theorem, stating that force is proportional to acceleration, does not come immediately from experience; logically considered, it is a free statement. It comes from the intuitively acquired knowledge that the phenomena of motion can be easily understood if acceleration is regarded as the fundamental phenomenon whose causes are sought. That this is not obvious in itself – to be precise, that it is not necessary – can be seen by looking at the history of mechanics before Galileo. The logical arbitrariness of this point of view is revealed by the fact that the general theory of relativity[6] has found it necessary to modify it.

Not only are fundamental laws the result of an act of imagination that can not be controlled, but so are their ingredients, the ideas derived from those laws. Thus, the concept of acceleration was in itself an act of free creation of the mind which, even if supported by the observation of the motion of solid bodies, assumes as a precondition nothing less than the infinitesimal calculus.

It follows from here that fundamental laws can be refuted not only by showing that the consequences attributed to them are wrong, inexact, or not generally applicable, but also can be refuted by showing that the concepts introduced for them do not suit the observed facts.

In this respect the history of modern theoretical physics offers beautiful examples. In the kinetic theory of heat, temperature is an elementary concept that stands out in a discussion on fundamental relations in that science. The development of thermodynamics showed that, in a body isolated from exchanges with its surroundings for any length of time, energy fluctuates permanently around a fixed average value; the smaller the portion of the body considered, the larger the fluctuations. If we observe parts that are sufficiently smaller, a precise distinction between its thermal and mechanical energy loses its meaning. The apparent incongruence of all these ideas is dispelled if we consider microscopically observable motions, such as those of very small particles suspended in liquids, as in the case of Brownian motion.[7]

---

[5] The text from here to footnote 6, is translated from German in *La Vida Literaria* (see fig. 6 in the Appendix).
[6] The text from here to footnote 7, is translated from Spanish in *La Vida Literaria.*

The process of progress in theoretical science finds its expression not only in the fact that the relations expressed by elementary laws are replaced by others that are more precise, but also in the circumstance that elementary concepts that are associated with the most immediate perceptions of reality need to be replaced by newer ones, better suited to the complex data provided by experience.

Will this development ever end?

We, contemporary physicists, no longer believe so. For us, any theory is true only in the sense in which a parable can be true.

However, if neither in this sense can we penetrate the ultimate truths, we are nonetheless left with the joyful awareness that each and every generation of researchers advances more profoundly toward the knowledge of what is true and real compared to that of its predecessors. In this sense, we wish to take joy from the work of our forerunners, and put forth our best effort on our part, while we place confidence in the strength of those who will come after us.

---

[7] The text from here to the end is translated from the German source reproduced in J. A. Stargardt's catalog 683, 188 (see fig. 4 in the Appendix).

## Appendix

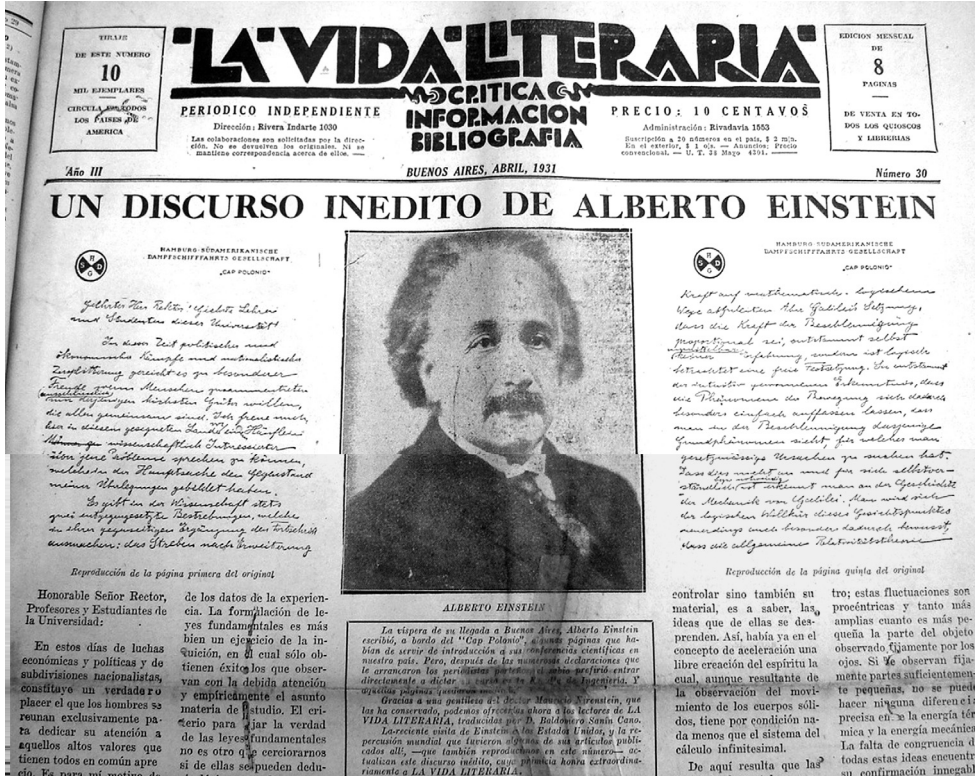

**Fig. 1.** Front page of the April issue of *La Vida Literaria*, Buenos Aires 1931. The Spanish translation of Einstein's lecture is visible below the photograph of Einstein flanked on the left and right side by the German text of the *discurso inédito* in the author's handwriting.

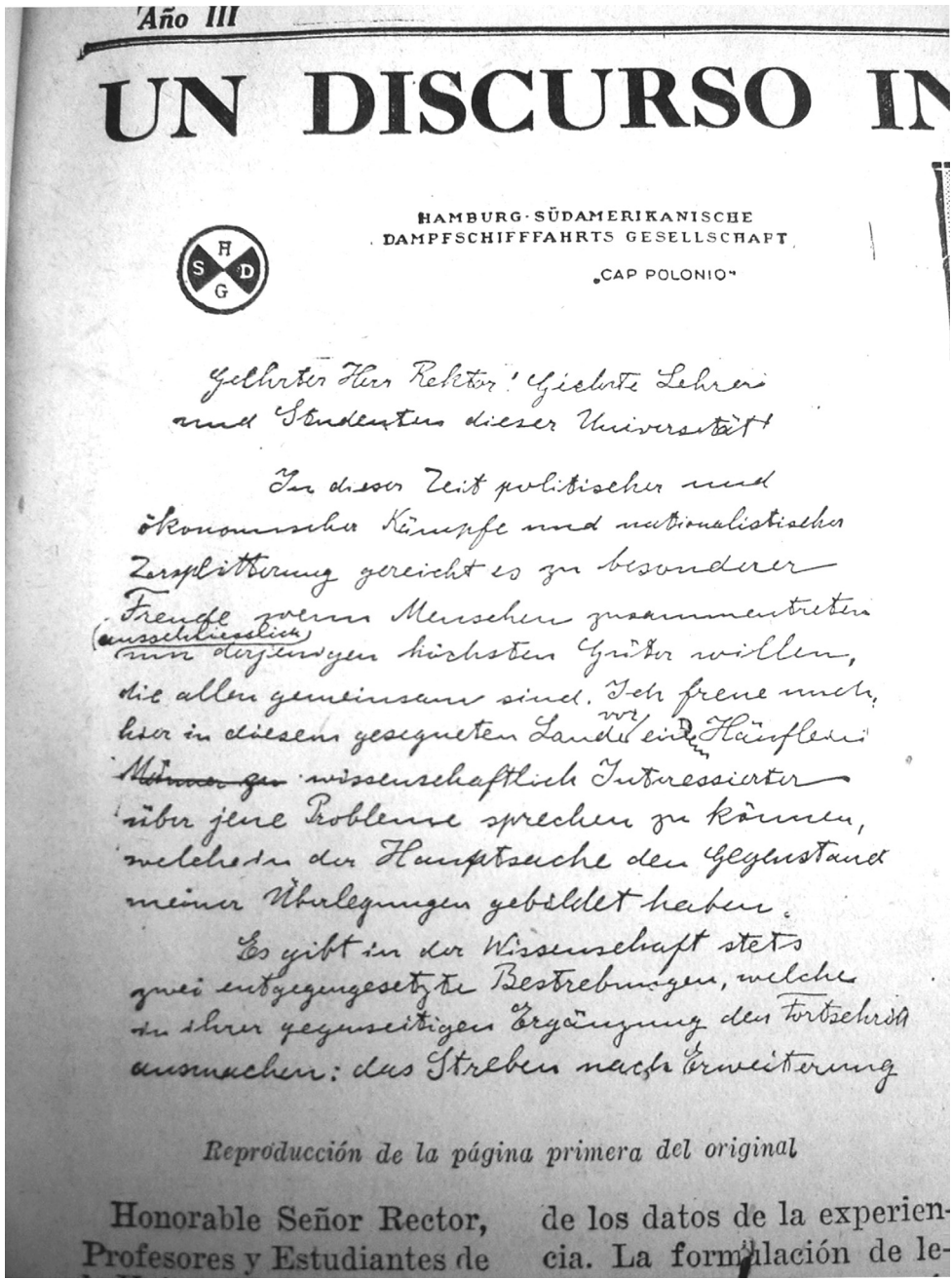

Reproducción de la página primera del original

**Fig. 2.** Detail of fig. 1, showing a reproduction of the first page of the original manuscript for the 1925 lecture.

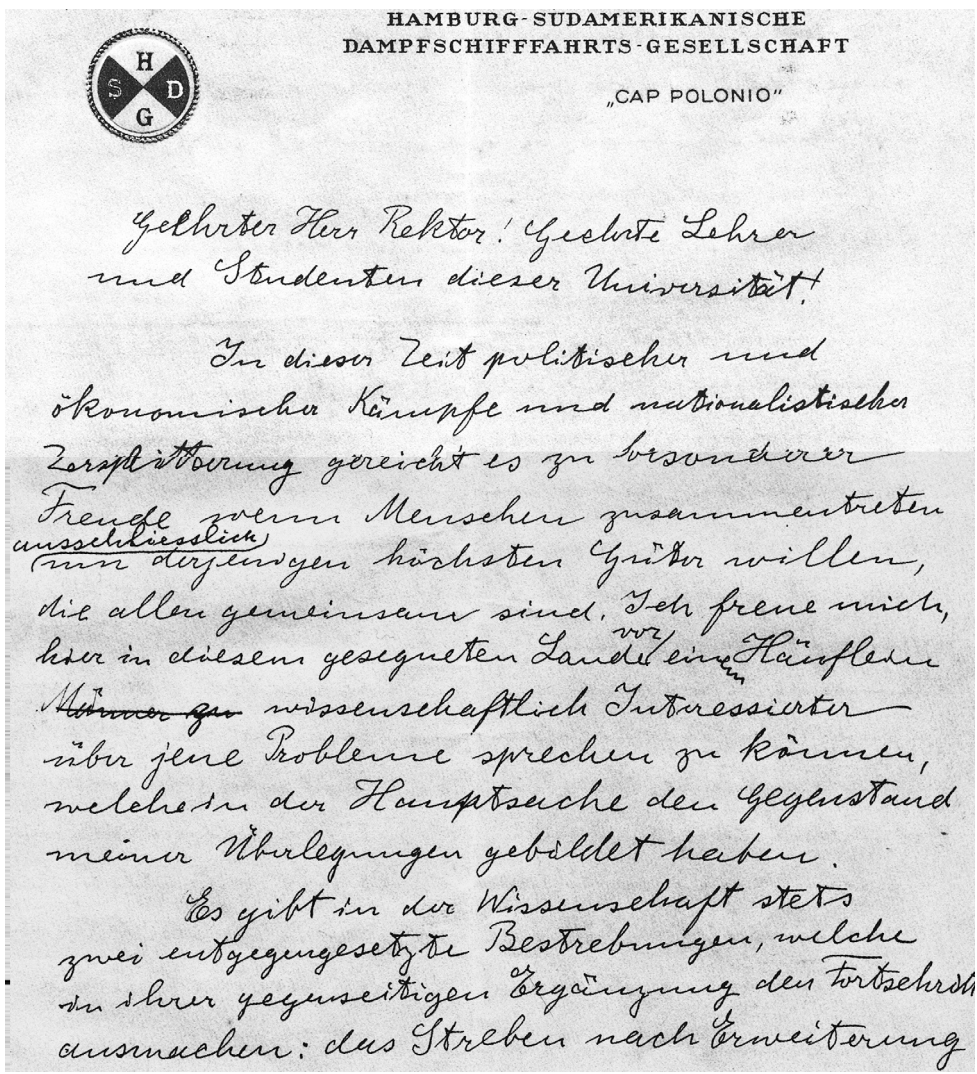

**Fig. 3.** The first page of Einstein's handwritten manuscript, reproduced in J. A. Stargardt's catalogs 615 (1978), 117, and 683 (2006), 188.

425 EINSTEIN, Albert, Physiker, Nobelpreisträger; Begründer der Relativitätstheorie, 1879–1955. Eigenh. Manuskript. (Frühjahr 1925.) 7¾ S. 4°, auf Briefbogen des Hapag-Schiffes „Cap Polonio". Mit eigenh. Streichungen und Zusätzen. Stellenweise leicht gebräunt.

(16.000.—)

Entwurf eines Vortrags über die Grundlagen der modernen Physik, den Einstein im März 1925 (in veränderter Fassung) an der Universität von Buenos Aires gehalten hat.

Nach den Einleitungsworten kommt Einstein auf die Dualität von Induktion und Deduktion in der Physik zu sprechen. Sein eigenes Streben sei weniger auf Bereicherung des Einzelwissens als auf systematische Einheit der Erkenntnis gerichtet.

„... *In dieser Zeit politischer und ökonomischer Kämpfe und nationalistischer Zersplitterung gereicht es zu besonderer Freude wenn Menschen zusammentreten ausschliesslich um derjenigen höchsten Güter willen, die allen gemeinsam sind. Ich freue mich, hier in diesem gesegneten Land vor einem Häuflein wissenschaftlich Interessierter über jene Probleme sprechen zu können, welche in der Hauptsache den Gegenstand meiner Überlegungen gebildet haben.*

*Es gibt in der Wissenschaft stets zwei entgegengesetzte Bestrebungen, welche in ihrer gegenseitigen Ergänzung den Fortschritt ausmachen: das Streben nach Erweiterung und Bereicherung unseres Einzel-Wissens, und das Streben nach systematischer Einheit der Erkenntnis. Da meine Arbeit stets dem letzteren Ziele gegolten hat, so will ich hier genauere Betrachtungen darüber anstellen.*

*Die Wissenschaft sucht das Geschehen auf möglichst wenige hypotetische Gesetze zurückzuführen, aus welchen sich die Relationen zwischen den beobachtbaren Thatsachen deduktiv, d. h. auf rein logischem Wege, ableiten lassen. Gewöhnlich nennt man die Physik eine empirische Wissenschaft und glaubt wohl, dass deren Fundamentalgesetze aus Experimenten abgeleitet seien, zum Unterschiede etwa von der spekulativen Philosophie. In Wahrheit ist aber die Beziehung der fundamentalen Gesetze zu den Erfahrungsdaten keine so einfache. Es gibt nämlich keine wissenschaftliche Methode, um induktiv die Fundamentalgesetze aus den Daten der Erfahrung abzuleiten. Die Aufstellung eines Fundamentalgesetzes ist vielmehr ein Akt der Intuition, der allerdings nur demjenigen gelingen kann, der das in Betracht kommende Gebiet empirisch genügend gut überschaut. Kriterium für die Wahrheit der Fundamentalgesetze ist allein dies, dass aus ihnen die empirischen Beziehungen zwischen den beobachtbaren Dingen bzw. Ereignissen logisch gefolgert werden können. Die Fundamentalgesetze können also wohl endgültig widerlegt, nie aber endgültig als richtig erwiesen werden; denn stets muss es als möglich in Betracht gezogen werden, dass ein Phänomen gefunden werde, das logischen Folgerungen aus den Fundamentalgesetzen widerstreitet. Die Erfahrung ist also wohl Richterin aber nicht eigentlich Erzeugerin der Fundamentalgesetze. Der Übergang von den Erfahrungsthatsachen zum Fundamentalgesetz bedarf stets eines freien schöpferischen Aktes der Phantasie, einer Schaffung von Begriffen und Relationen, ohne dass es möglich wäre diesen Akt durch eine zwangläufige Methode zu ersetzen ...*

*Der Prozess des Fortschrittes der theoeretischen Wissenschaft äussert sich ... nicht nur darin, dass die in den Elementargesetzen ausgedrückten Relationen durch genauere ersetzt werden, sondern vor allem auch darin, dass die elementaren Begriffe, welche den letzten Realitäten entsprechen sollen, durch neue, dem Komplex der Erfahrung mehr adäquate ersetzt werden.*

*Gibt es ein Ende in dieser Entwicklung? Wir heutigen Physiker glauben es nicht mehr. Alle Theorie hat für uns Wahrheit nur in dem Sinne, in dem ein Gleichnis Wahrheit enthalten kann.*

*Wenn es aber auch in diesem Sinne kein Vordringen zu letzten Wahrheiten geben mag, so bleibt uns doch das frohe Bewusstsein, dass jede Generation der Forschenden in der Erkenntniss des Wahren und des Realen tiefer vorschreitet als ihre Vorgängerin. In diesem Sinne wollen wir uns unserer Väter freuen, wollen wir selbst weiter streben und auf die Kraft der Späteren vertrauen."*

Am Schluß eine Echtheitsbestätigung seines Sohnes Hans Albert Einstein.

**Fig. 4.** Fragment from Einstein's lecture printed in J. A. Stargardt's catalog 683 (2006), 188.

389  EINSTEIN, Albert, Physiker, Nobelpreisträger; Schöpfer der Relativitätstheorie, 1879–1955.
Eigenhändiges Manuskript. 8 S. 4°. Stellenweise leicht gebräunt.                    (8000.—)

Entwurf eines Vortrags über die G r u n d l a g e n   d e r   m o d e r n e n   P h y s i k , den Einstein im
März 1925 (in veränderter Fassung) an der Universität von B u e n o s   A i r e s  gehalten hat. Das
Manuskript ist auf Briefbogen des Hapag-Schiffes „Cap Polonio" geschrieben.

Nach den Einleitungsworten kommt Einstein auf die Dualität von Induktion und Deduktion in der
Physik zu sprechen. Sein eigenes Streben sei weniger auf Bereicherung des Einzelwissens als auf syste-
matische Einheit der Erkenntnis gerichtet.

„*. . . Gewöhnlich nennt man die Physik eine empirische Wissenschaft und glaubt wohl, dass deren
Fundamentalgesetze aus Experimenten abgeleitet seien . . . In Wahrheit ist aber die Beziehung der
fundamentalen Gesetze zu den Erfahrungsdaten keine so einfache. Es gibt nämlich keine wissenschaft-
liche Methode, um induktiv die Fundamentalgesetze aus den Daten der Erfahrung abzuleiten. Die Auf-
stellung eines Fundamentalgesetzes ist vielmehr ein Akt der Intuition, der allerdings nur demjenigen
gelingen kann, der das in Betracht kommende Gebiet empirisch genügend gut überschaut. Kriterium
für die Wahrheit der Fundamentalgesetze ist allein dies, dass aus ihnen die empirischen Beziehungen
zwischen den beobachtbaren Dingen bzw. Ereignissen logisch gefolgert werden können. Die Fundamen-
talgesetze können also wohl endgültig widerlegt, nie aber endgültig als richtig erwiesen werden; denn
stets muss es als möglich in Betracht gezogen werden, dass ein Phänomen gefunden werde, das logi-
schen Folgerungen aus den Fundamentalgesetzen widerstreitet . . .*" usw.

**Fig. 5.** Fragment from Einstein's lecture printed in J. A. Stargardt's catalog 615 (1978), 117.

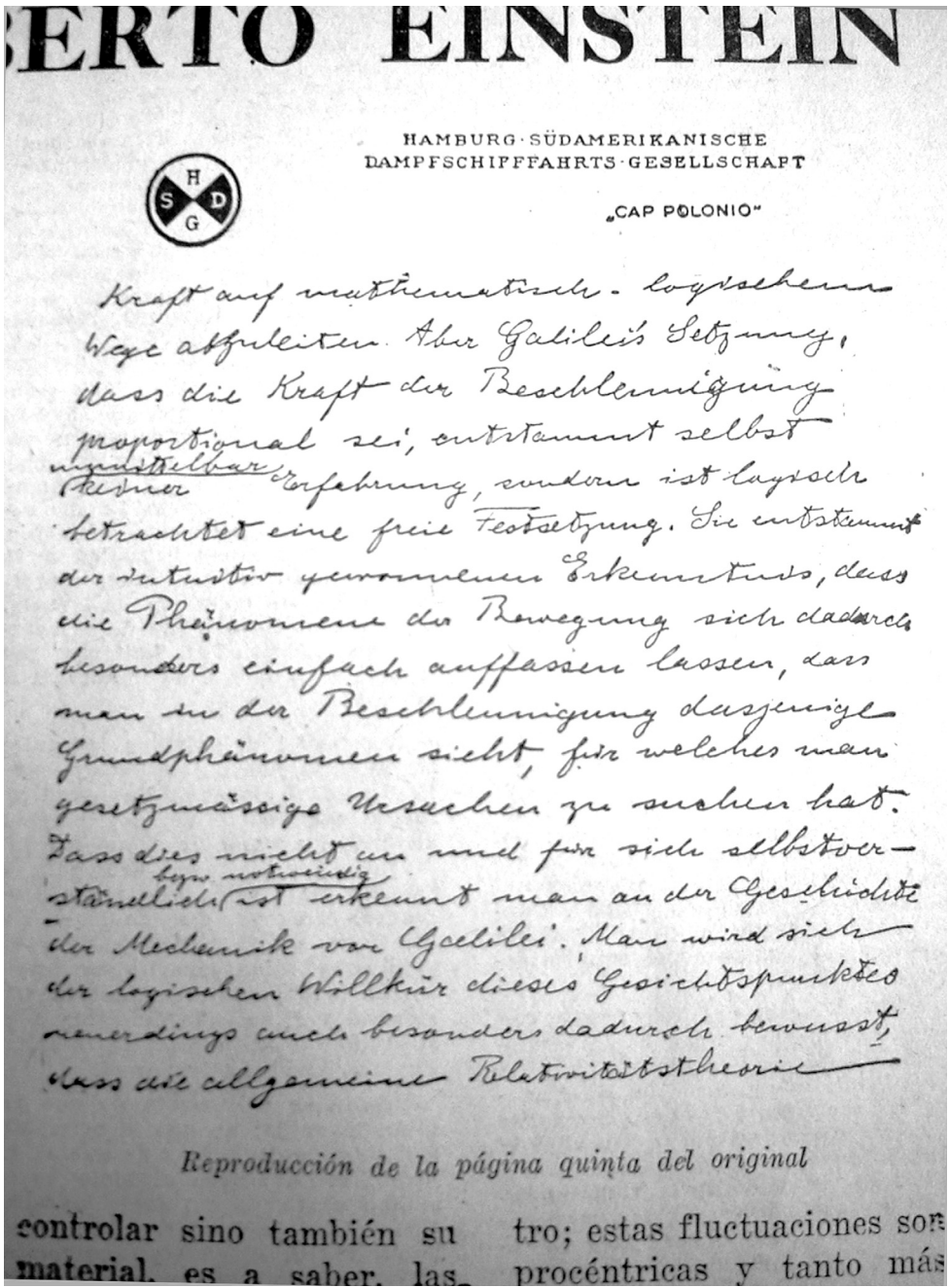

**Fig. 6.** Detail of fig. 1, showing a reproduction of page 5 of Einstein's handwritten manuscript.